\documentclass[a4paper]{article}
\usepackage{graphicx}

\def\be{\begin{equation}}
\def\ee{\end{equation}}

\def\d{^\circ}

\title{Stellar complexes in M33}
\author{G.R.Ivanov}
\begin{document}
\maketitle 
\abstract{A method for identification of stellar complexes in M\,33 is applied.  Several OB associations form a stellar complex with mean size of 0.3 - 1 kpc. We apply a correlation technique to compare different stellar populations in M\,33. Our results confirm the existence of a strong correlation between OB stars, H\,{  II} regions and WR stars, which trace the regions of massive star formation. There is a good correlation between red supergiants (RSGs) and WR stars in the spiral arms of M\,33. This can be expected since the progenitors of WR stars are massive OB stars or RSGs with masses $\rm M \geq 20 M_{\odot}$. The massive RSGs and WR stars probably originate from nearby sites of star formation. We consider this fact as a ground for identification of stellar complexes in M\,33. The presented method for identification of stellar complexes can be applied to other nearby galaxies.}

\section{Introduction}
\begin{table}
Center's coordinates (nucleus):\\
$R.A. (2000) = 23.462042 \d$ \\
$Dec. (2000) = +30.6601944\d$ \\
Distance modulus: m-M = 24.52 \\
\caption{General information on M33 galaxy}
\end{table}
M33 is a nearby galaxy of Sc type with a suitable inclination between galactic disk and the plane of sky for studding the stellar association, which form star compexes. In this galaxy Humpheys and Sandage (1980) isolated 143 associations with a mean diameter 200 pc.  Efremov (1995) considers the star complex as a slellar group of young stars form together by fragmentation of a dense molecular cloud, which are initial structures owing to large-scale gravitational instability. They consist of objects with total masses about $ 10^7 M_{\odot}$. The aim of the present paper is to propose a method for identification star complexes in M33 using observational data of HII regions which are exited by massive OB stars embedded in them. On the other side, WR stars are objects, which are physically associated with HII regions. We suppose that OB stars WR stars, HII regions, Red supergiants (RSGs), Cepheids and supernova remats (SNR)  indicate the star complexes as regions of massive star formation. There was evidence of large-scale groupings of RSGs and Cepheids not within OB associations but beyond them through within vast groups (Efremov, 1995). So the possible stellar populations of star complexes may be also RSGs and Cepheids.(RSGs in M33 were selected by Vassilev et al. (2002), observed in JHK passbands and published in 2MASS survey was used. The sample of RSGs was defined by the limit $ K>13$ mag. In this criterion was removed the brighter stars because they should belong to Milky Way background.  The sample of RSGs contains 1650 stars selected by criterion $J-K>1.1$ .which are suitable for the purpose of present paper. Observational data for coordinates of 905 Cepheids in the central region o M33 are from Mochejska et al. (2001). However the data of latter paper do not cover the total area of M33 because of the observational data of Sandage (1983) are also used in the present study. The gas component and the stellar distribution of M33 are well studied. Catalogues of HII regions have been published by and Courtes et al. (1987).  Macri et al. (2001) present a catalogue of 319 OB stars in the cental region of M33.  Ivanov et al. (1993) (hereafter IFM) presented a catalog of 2112 OB stars. Massey and Conti (1983) and Massey et al. (1987) have presented a list of the Wolf-Rayet stars. Massey et al. (1995). Nowadays the number of WR stars in M33 amounts 168. Gordon et al. (1998) have inetified 98 supernova remants (SNR).
\section{Correlation technique}
In the present study we combine the above mentioned surveys of OB stars, H\,{  II} regions and WR stars in order to compare the distributions of these objects with the population of RSGs and Cepheids. We use also a correlation technique
for comparison of stellar populations, proposed by Ivanov (1998). Let N1 stars of one population in M\,33 have
surface density $\delta_1$ while another population of N2 stars has a surface density $\delta_2$. The two-dimensional
angular distance between the stars of the $k$-th stellar couple is $\rm d_k$ as defined in the Appendix of Ivanov (1998).
Supposing a random distribution of the stellar populations in the galaxy, the distance between the two stars of the $k$-th
couple is exactly $\rm d_k$. Then the probability:
\be \label{1}
\rm P_ {12}(k) =  [(1- exp(-\pi d_k^2\delta_1)] [(1-exp(-\pi d_k^ 2\delta_2)],
\ee
is a measure for associated stars (see Ivanov 1998). The actually associated stars of two different populations form
couples with $\rm d_k \rightarrow0$ and consequently $\rm P_{12}(k)\rightarrow 0$. The couples of foreground stars have
a bigger neighbor distances $\rm d_k$ and $\rm P_{12}(k) \rightarrow 1$. The quantity $\rm P_{12}(k)$ gives the probability
to find one star of population "1" and another star of population "2" within a radius equal to  $\rm d_k$ in case both the
populations are randomly distributed. Probabilities $\rm P_{12}(k)\approx 0$ can be used as a good criterion for
associated couples. The presented criterion defines the upper and the lower limit of probability $\rm P_{min}$ and
$\rm P_{max}$ which can be obtained from observational data. If $\rm P_ {12}(k) < P_{min}$, a couple of associated stars is
selected, while foreground couples are defined by $\rm P_ {12}(k) > P_{max}$. Further we obtain $\rm P_{min}=0.05$
and $\rm P_{max}= 0.95$. These values are generally accepted in the statistics. In case when the associated stars are
selected by criterion  $\rm P_{12}(k) < 0.05$, the number of associated couples is indicated with N5. A stronger criterion
for selecting associated stars and foreground couples is imposed, if the individual probabilities of the couples
$\rm P_ {12}(k) < 0.01$ and $\rm P_{12}(k)> 0.99$, respectively. In this case, the number of associated couples is denoted
with N1. Let the number of foreground couples is $N_fgr$. A simple way to evaluate the correlation between two populations
is to obtain the percentage of associated objects:
\be \label {2}
\rm R5 = N5 / N; \hspace {5mm} ;R1 = N1 / N.
\ee
The ratios R1 and R5 are very suitable measures for correlation between the stellar populations. If all stars of two
populations are associated with each other, then $\rm R5=1$ or $\rm R1 = 1$. In the opposite case (no associated stars
at all), $\rm R5 =0$ or $\rm R1 = 0$. The ratios, given by Eq. \ref{2}, are analogous to the conventional coefficient of
correlation in the statistics. Another way to evaluate the correlation between two stellar populations is to calculate the
ratio of the number of associated objects to the expected number from random distribution:
\be \label {3}
\rm RN5= N5 /N_{fgr}\hspace{5mm} ;RN1 = N1/N_{fgr}.
\ee
 \begin{table}
\caption{Correlation parameters between stellar populations in M\,33}
\begin{center}
\begin{tabular}{ | c | c c c c c c c c|}
\hline
Correlation parameter &  OB-HII&  HII-WR  & RSG - WR & WR - OB  & Ceph- WR& RSG-HII & RSG-Ceph &RSG-SNR\\
\hline
R5 & 0.52  &      0.76  &   0.70  &   0.89 & 0.25 & 0.09& 0.97 &0.98\\
\hline
R1 & 0.38   &      0.54  &   0.43  &  0.74 & 0.14 & 0.04& 0.83 &0.83\\
\hline
RN5 & 2.2  &     7.4  &   16.3 &   91  & 0.74 & 0.12& 135& 98\\
\hline
RN1 & 1.7  &      5.5  &   10  &   26  & 0.40& 0.06 & 114& 96\\
\hline
Number of couples  &  748  &  140   & 140   & 140 & 140 &748 & 147&98\\
\hline
\end{tabular}
\end{center}
The contents of the table are as follows:

Column 1 gives the name of correlation parameters.\\
Column 2 gives the correlation parameters between OB stars and H\,{  II} regions.\\
Column 3 gives the correlation parameters between H\,{  II} regions WR stars.\\
Column 4 gives the correlation parameters between RSGs and WR stars.\\
Column 5 gives the correlation parameters between OB stars  and WR stars.\\
Column 6 gives the correlation parameters between Cepheids  and WR stars.\\
Column 7 gives the correlation parameters between RSGs and H\,{  II} regions.\\
Column 8 gives the correlation parameters between RSGs and Cepheids \\
Column 9 gives the correlation parameters between RSGs and SNR \\

\end{table}

\section{Correlation between stellar populations in M33}

The data presented in Table 2 can be interpreted as an evidence of weak correlation between SNR, OB  stars HII regions.   The tight correlation between SNRs and RSGs is expected since the the  RSGs evolve to SNRs. Hence a part of RSGs used in the present study should a strong correlate with SNRs regions. We suppose that a fraction of the OB stars which excite the HII regions and belong to stellar associations is not detected up to  because of large extinction in the optical part of spectrum (UBV photometry) of regions for stellar associations and partly the extenuation in the spectral line $H_{\alpha}$ for HII regions. There is no correlation between RSGs and  HII regions and between Cepheids and WR stars. However take into account that in our paper  Vassilev el al. were selected RSGs with masses 12 -20 $M_{\odot}$  It is well known that SN have masses in the range of RSGs, selected by Vassilev et al. (2002). The correlations between stellar populations in M33 are given in Table 2.  This can be expected because the progenitors of SN stars are RSGs which  have masses $\rm M \geq 20 M_{\odot}$. On the other hand, the massive RSGs as well supenona  stars must originate from nearby sites of star formation.  If a star complex is a huge group of clusters, OB associations, HII regions and high luminosity stars, accepted by Efremov(1995), we have to surch the star complexes in the regions of physically associated objects with high surface density comparing to surrounding objects. 

\section{Star complexes in M33}
The cluttering  method for the identification the stellar groups are described in the paper of Ivanov (1996). The criterion proposes that the objects will be assigned  to the one and the same group if they have statistically peak of surface density above the mean level of the neighbor objects. In other words the surface density is the main property that can isolate the star complex from the surrounding objects. The data of Table 2 can be considered that there are real physical associated between the objects of OB stars, HII regions and WR stars because of we identify the star complexes in this section as regions of physically associated objects of  OB stars, HII regions and WR stars with a peak density of above 5 times above the surrounding density of these objects. When  the  site of density peak was defined, we take into account the surface density of additional objects as RSGs and Cepheids in the boudoirs of  star complexes. The  Cepheids and RSGs do not show a considerable concentration to stellar group in M33. Beside RSGs do not correlate with the main counterpart of star complexes OB stars and HII regions. For this reason we do not expect many RSGs and SNRs to fall in the same star complexes. However we fount a lot SNRs and RSGs in some star complexes while they are in  deficit in other star complexes. We suppose that the presence of these objects in some star complexes gives evidence for the connection between the RSGs and RSGs through the stellar evolution process in parent molecule cloud.

\section{Discussion}
The boundaries of star complexes, using  the peaks of surface density in OB , WR stars and HII regions distribution are delineated in Fig. 1. The coordinates of the center of complexes and surface density of the objects  within a star complex are given in Table 3. The complexes with numbers  21, 23, 28, 29, 31, 33, 34, 36, 38, 40, 41a, 43 are located around  the center of M33 (outlined by a large plus sign). They contain many OB stars and bright HII objects. There is observational evidence that bright HII regions exhibit a strong concentration of OB stars toward their centers. The coincidence of OB stars and HII region in the same star complex is a good indicator for age estimation of the complex. We suppose these star complexes indicate the extended central region of M33 with the youngest objects of the galaxy. At the same time, these complexes contain many WR stars. The presence of WR stars in the central part of M33 indicates their star complexes as regions of massive star formation. On the other hand, the star complexes with the numbers 4, 5, 8, 18, 25, 29, 33, 37, 41, 46, and 47 outline well the two main spiral arms. Their stellar population differs from that of star complexes in the central region of the galaxy M33. \\ 
\newpage
There is a good correlation between WR stars and RSGs in these complexes. This fact is explained by Georgiev \& Ivanov (1997) who studied the distribution of RSGs of IFM and WR stars as a function of galactocentric distance in M33. They suggested, by comparing both distributions, that RSGs with masses higher than 30 $\rm M_{\odot}$ would evolve to WR stars, whereas the less massive should spend some part of their lives as RSGs. If our sample of RSGs have masses below this limit,  in the range of $12-20 M_{\odot}$, we expect to find WR stars and their progenitors RSGs at the same sites in the galaxy. The tight correlation between RSG and WR stars in Table 2 confirms their suggestion. This correlation speaks about the disposition of the two populations on the same or nearby sites in the galaxy while the stars dispose on remote regions in the galaxy have a negligible influence of our correlation parameters. The sample of RSGs confines all regions of the galaxy. Then in the metal rich regions of star complexes, some RSGs would evolve to WR stars and may disappear. But in a less metal rich regions the progenitors may be in stage of RSGs. So, the correlation between  WR tars and RSGs occur in the region of chemical abundance as spiral arm regions. This fact explains because RSGs may exist or lack in some star complexes in Fig. 1. We can conclude from the stellar populations of star complexes in Table 3 that predictions of Maeder et al. (1980) for the influence of metallicity over the star's lifetime are confirmed by observations. The  chemical composition in star complexes of M33 is better defined than in other galaxies. However, if should take into account the local imhomogenities in the abundance of heavy elements which may explain the variegated the  number of WR stars and RSGs from various star complexes.\\
We combine the distributions of WR stars and RSGs in star complexes. Therefore, if some class of RSGs were progenitors of WR stars we would have to compare the distribution of WR stars with different samples of RSGs. So, the distribution of WR stars should be compared with RSGs with different masses. Then we can obtain the sample of RSGs, which are probable progenitors of WR stars. Since, the distribution of WR stars is fixed we  have a criterion to evaluate the masses of RSGs in spiral arm complexes. The value of magnitude at which the distribution of the classes of the progenitors and descendants coincide using data of Table 3 will show the mass of RSGs which evolve to WR stars. Then, take into account data of Table 3 we conclude that RSGs in M33 with masses $15-20 M_{\odot}$ evolve to WR stars.
\begin{table}
\renewcommand{\arraystretch}{0.9}
\caption{Star complexes in M33}
\begin{center}\tabcolsep=3pt\small
\begin{tabular}{ | c | c c c c c c c c c c c c c c|}
\hline
1 & 2 & 3 & 4 & 5 & 6 & 7 & 8 & 9 & 10 & 11 & 12 &13&14&15\\
No &  R.A.(2000)&  Dec.(2000)  & $OB_N$ &$ OB_F$  &  $HII_N$ &$ HII_F$  & $WR_N$ & $ WR_F$  &  $RSG_N$ & $ RSG_F$  &  $Ceph_N$ & $ Ceph_F$ & $SN_N $ & $SN_F$\\
\hline
 1& 23.4297& 30.4200 & 94 & 5.6 &  65&  6.7&   4 & 5.6&     9 &   4.1&    14 * & 6.6 &0 & 0.0\\
 2 &23.4300 &30.3600 &129&  6.1&  51&  6.5&   4 &  5.6&  32&  7.4 &    9 * & 13.0 & 2&0.0 \\
 3 &23.4303 &30.6450 &469  &6.8 &126 & 5.6&   23 & 7.4 & 47& 4.6 &  70 &  5.5 & 4& 4.9 \\
 4 &23.4332 &30.5850 &438  &6.8 &116  &6.6&  40  &6.8  & 50&  6.5  & 89&  4.7 &4& 4. 4\\
 5& 23.4332 &30.6400 &492  &7.1 &129  &6.3&  22  &8.5  & 52&  4.3  &73&  5.7 &4& 4.0\\
 6& 23.4340 &30.3995 &104  &7.5  &70  &7.2  &  3  &9.9  &12 & 7.0   &  12 *  & 5.2 &1&1.8\\
 7&23.4370 &30.5700 &461  &8.1 &126  &6.6 & 47  &7.7  &67  & 4.3  & 96  &4.2 &2&0.0\\
 8& 23.4375& 30.3817&128  &8.7 & 61  &8.8   & 4  &9.4 &  22 & 6.9  &  18 *  & 9.0 &7&3.5\\
 9& 23.4376& 30.4457 & 97  &8.5  &54  &9.2   &6 & 10.5&   6  & 4.5    &4  & 2.6 &0&0.0\\
10& 23.4380& 30.6470 &500 & 9.3 &138&  6.9  &24 & 9.8&  51& 5.0  & 70 & 5.9 &4&4.9\\
11 &23.4387 &30.5020 &206 &7.0 &  71 & 6.9 &19  &6.1 & 21 & 3.8 &  70  & 5.1&2&5.7\\
12 & 23.4389 & 30.7450 & 308 & 6.3 & 85 & 7.8 & 20 & 11.1 & 40 & 7.5 & 84 & 6.0 &3&3.8\\
13  & 23.4391 & 30.8870 &140 &10.4 & 60 & 6.8 &  0  & 0.0  &16  &6.9   &11*  & 6.7 &1&7.2\\
14  & 23.4402 &30.7950 &267  &6.7  &93  &7.8   & 8   &6.8  &14  &4.4  &55  &6.3 &0&0.0\\
15  &23.4409  &30.4400  &97  &7.6  & 55  &9.4    &5   &7.7   & 5  &4.6   & 9 *  & 11.2 &0&0.0\\
16  &23.4422  &30.6467  &    5  &10.4& 138  &7.7  &25  &9.9  &52  &5.2  &73  &6.0 &4&4.7\\
17  &23.4422  &30.6800  &478 &6.6 &133  &7.3   &38   &9.5  &54   &5.3   &47  & 5.7 &8&4.9\\
18  &23.4428  &30.5250  & 315 &8.3 &105  &9.4  &30   &8.3  &40 &6.3  &104  &4.6 &4&5.9\\
19  &23.4429  &30.5560  &451& 7.8 &124  &7.5   &48   &7.0  & 57 &4.4 &112  &4.3 &6&4.2\\
20  &23.4438  &30.8020 &241  &6.9  &96  &8.1      &7   &9.7  &13  &4.6  &54&6.2&0 &0.0\\
21  &23.4447  &30.6400  &513  &9.8 &134  &8.2   &23  &10.7  &54  &4.9  &77  &6.0 &3&3.8\\
22  &23.4453  &30.8667  &142  &9.6   &60  &8.8   & 0  & 0.0   &19  &8.4   & 12 *  &9.1 &1&7.4\\
23  &23.4453  &30.6050  &461  &6.2  &125 &7.4  & 34  &6.1   &45  &6.7  &78  &4.8&3&6.8\\
24  &23.4455  &30.7150  &309  &6.9 &112  &6.6  & 38  &6.2  &45  &6.2  &75  &6.1&5&4.4\\
25  &23.4457  &30.3775  &130   &9.0  &57  &9.1   &4   &8.1   &24  &6.2   & 11 *  &9.1 &1&1.8\\
26  &23.4458  &30.3450  &110  &8.6   &42  &6.0   & 3 &20.2  &29  &5.5   &8 *  &25.1 &1&8.8\\
27 &23.4459  &30.7560  &339  &8.4  &93  &8.0  &21 &11.9  &36  &6.5  &99  &5.2&2&5.8\\
28 &23.4459  &30.6895  &389  &6.8 &124  &7.2  &34 &12.1 &55  &5.0  &56  &5.0 &8&3.9\\
28a& 23.4465 &30.7395 &307 &7.1  &88  &8.3  &22 & 9.4  &39  &7.6  &90  &6.9&3&4.8\\
29  &23.4468  &30.5960 &438  &7.5 &128  &6.9  &37  &7.9  &46  &6.8  &82  &4.5 &3&7.5\\
30 &23.4480 &30.8767 &142 &12.6  &63  &8.1   & 0  &0.0  &18  &8.1   & 13 *  &6.8 &1&4.9\\
31 &23.4483 &30.6415 &523 &10.2 &133  &8.6  &23 &11.0  &58  &4.9  &75  &6.4 &3&4.3\\
32 &23.4499 &30.7457 &322  &7.8  &92  &8.3  &21 &12.4  &38  &7.4  &92  &6.6 &2&8.2\\
33 &23.4502 &30.5540 &441  &7.8 &127  &7.3  &47  &7.1  &53  &4.3  &99  &4.9&9&3.7\\
34 &23.4511 &30.6580 &522 &10.6 &146  &8.1  &36  &6.4  &56  &6.9  &70  &6.0 &5&5.0\\
35 &23.4511 &30.7750 &321 &10.3  &95  &7.3  &19  &6.8  &32  &4.7  &97  &5.2&1&5.1\\
36 &23.4512 &30.5986 &443  &6.9 &131  &6.8  &38  &7.0  &50  &5.7  &76  &4.9 &4&5.6\\
37 &23.4535 &30.7780 &320  &9.9  &98  &7.1  &19  &5.6  &34  &4.4  &93  &5.5&1&4.2\\
38 &23.4535 &30.6560 &515 &10.4 &143  &8.1  &36  &5.8  &54  &7.4  &77  &5.9 &4&5.6\\
38a &23.4539 &30.7795 &318 &9.6  &99  &7.0  &19  &5.6  &34  &4.4  &87  &5.8 &1&4.2\\
39 &23.4555 &30.8875 &143 &12.1  &61  &7.8   &0  &0.0  &15  &6.6   & 11 *  &7.8 &4&5.8\\
40 &23.4558 &30.6300 &532  &7.0 &135  &7.3  &25  &6.6  &51  &4.5  &91  &5.2&1&4.3\\
41a &23.4559 & 30.5595 &453 &7.2 &128  &6.5  &48  &7.9  &58  &4.1 &103  &4.6 &2&0.0\\
41 &23.4579 &30.7615 &338  &9.8  &97  &6.8  &22 &10.0  &41  &5.3 &109  &5.5 &4&3.6\\
42 &23.4600 &30.9449  &59 &10.3  &37  &8.9   &2  &0.0   &8  &8.9   & 8 *  &10.7 &8&4.4\\
43 &23.4604 &30.5965 &423  &6.1 &123  &6.0  &36  &6.3  &52  &4.9  &71  &5.4 &2 &7.9\\
44 &23.4607 &30.6802 &449  &6.2 &130  &6.3  &37  &6.9  &59  &6.3  &73  &4.4 &5&4.7\\
45 &23.4607 &31.0095  &25 &14.4  &20  &9.2   &0  &0.0   &4  &5.1   & 0  &0.0 &0&0.0\\
46 &23.4613 &30.7775 &317  &8.8  &94  &6.2  &18  &5.3  &36  &5.0  &99  &5.6 &2&4.8 \\
47 &23.4678 &30.6875 &360  &5.4 & 110  &5.4  &30  &6.3  &58  &5.4  &5  &4.2 &5&4.2\\
\hline
\end{tabular}
\end{center}
The contents of the table are as follows:\\
Column 1 gives a running number of star complexes according to increasing right ascensions.\\
Columns 2 and 3 give the right ascensions and declinations for equinox 2000.0. in degree\\
Columns from 4, 6, 8,10 and 12 and 14  give the number of populations within the star complex.\\
Columns from 5, 7, 9,11 and 13  and 15 give the density of stars within the star complex compared to background object.\\
 $*$ The number of Cepheids in column 12 denoted with asterisk $^*$ are based on data of Sandage (1983).
\end{table}
\begin{table}[h]
\caption{Regions of star complexes in M33}
\begin{center}
\begin{tabular}{ | c |  c | c |}
\hline
Region No &  Star complex No &  Remark \\
\hline
center & 21; 23; 28; 29; 31; 33; 34; 36; 38; 40; 41a; 43& OB + HII regions+WR + SNR\\
\hline
S1     & 4; 5; 8; 18; 25; 29; 33                                     & OB +WR + RSG + Ceph + SNR\\
\hline
N1      & 16; 17;24; 27; 29; 32 36; 37 ; 41; 46; 47           & OB + WR + RSG+ Ceph + SNR\\
\hline
North &13; 22; 30; 32; 42; 45                                       & Without WR \\
\hline
South  & 1; 2; 6; 15; 25; 26                                           & Without WR \\
\hline
\end{tabular}
\end{center}
The contents of the table are as follows:

Column 1 gives  Name of the regons\\

A column 2 gives the numbers of star complexes within the region.\\
Columns 3 give the stellar population of the region\\
\end{table}
Regions of star complexes may be considered as group of star complexes with violent star formation region.\\
{\bf Acknowledgements}
\rm This project was partially supported by grant (F-1302/2003) of the Bulgarian National 
Science Foundation. \\ \\
{\bf REFERENCES}\\
Courtes, G., Petet, H., Sivan, J-P., Dodonov, S., \& Petit, M., 1987, A\&A, {\bf 174}, 28\\ 
Efremov, Yu. N., AJ, 1995, {\bf 110}, 2757\\
Ivanov, G. R., 1996,  A\&A, {\bf 305}, 708\\
Ivanov, G. R., 1998,  A\&A, {\bf 337}, 39\\
Georgiev, L. \& Ivanov, G. R., 1997, RevMexAA, {\bf 33}, 117\\
Gordon, S. M., Kirshner, R. P., Long K. S., Blair, W. P., Duric N., \& Chris Smith, R., 1998, ApJS, {\bf 117}, 8\\
Ivanov G., Freedman W., \& Madore F., 1993, ApJS, {\bf 89}, 85 (IFM)\\
Macri, L. H., Stanek, K. Z., \& Sasselov, D. D., 2001, AJ, {\bf 121}, 861\\
Massey, P.  \& Conti, P. S., 1983, ApJ, {\bf 273}, 576\\
Massey, P., Conti, P. S., Moffat, A. F. J. \& Shara, M.M., 1987, PASP, {\bf 99},  816\\
Mochejska, B. J., Kaluzny, J., Stanek, K. Z., Sasselov, D. D., \& Szentgyorgyi, A. H., 2001, AJ, {\bf 121}, 2032\\
Vassilev,  O., Vassileva, L., Ivanov, G.R., \& Vassilev,  D., 2002, Publ. Astron. Obs. Belgrade, {\bf 73}, 257, \\
Sandage, A., 1983, AJ, {\bf 88},1108\\

\begin{figure}[h]
\begin{center}
\includegraphics[scale=0.7]{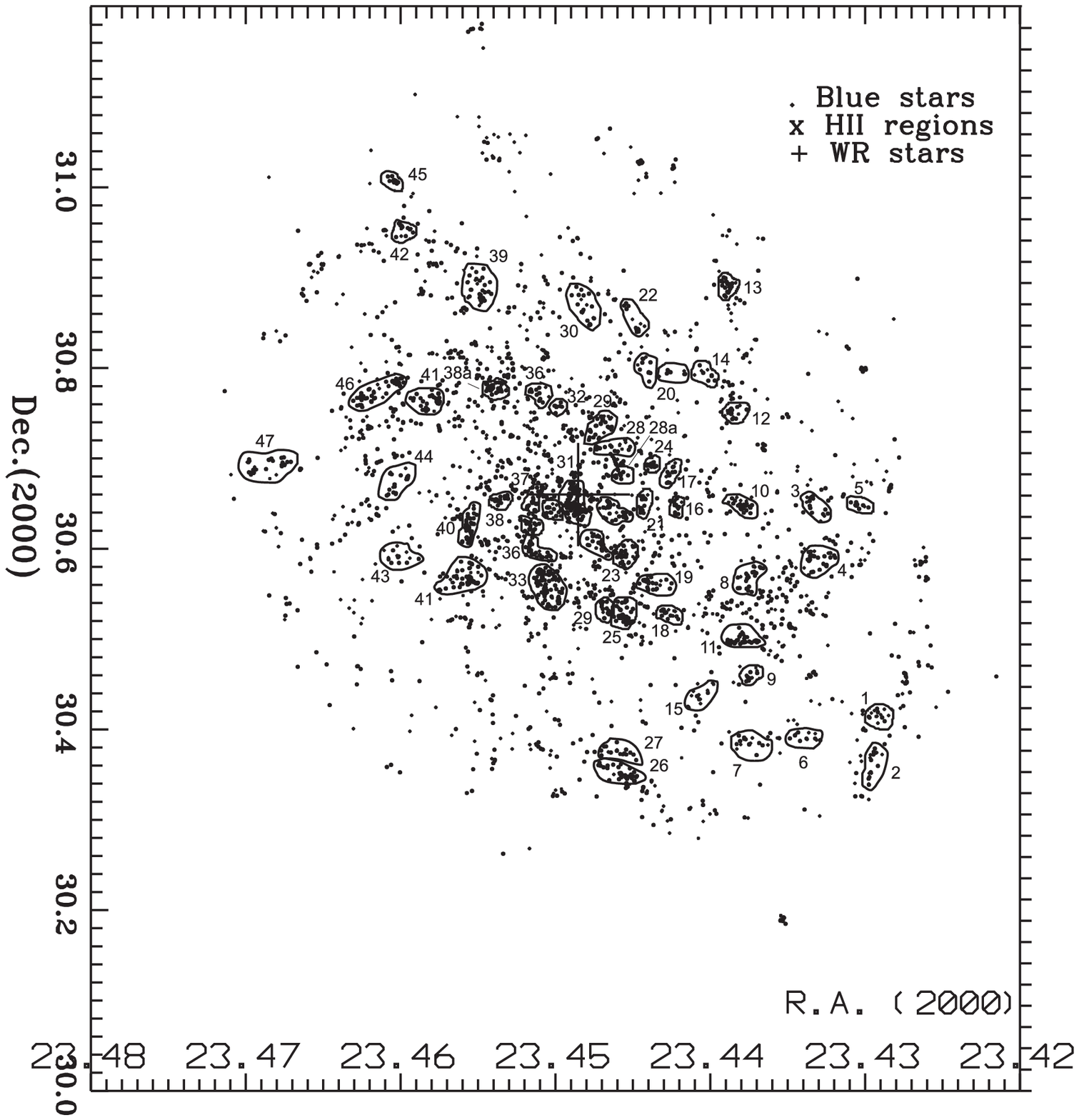}
\caption{The stellar complexes in M\,33.}
\end{center}
\end{figure}

\end{document}